\documentclass[copyright,creativecommons]{eptcs}

\usepackage{breakurl}             
\usepackage{alltt}
\usepackage{graphicx}
\usepackage{url}

\usepackage{amsmath,xcolor,moreverb,url,float,amssymb,listings,subfigure}

\title{Polychronous Interpretation of Synoptic, a Domain Specific Modeling Language for Embedded Flight-Software}
\author{L. Besnard, T. Gautier, J. Ouy, J.-P. Talpin, J.-P. Bodeveix, A. Cortier, \\ M. Pantel, M. Strecker, G. Garcia, A. Rugina, J. Buisson, F. Dagnat
}

\newcommand{\signal}{{\sc Signal}}

\newcommand{\spacify}{SPaCIFY}

\newcommand{\V}{{\cal V}}

\renewcommand{\S}{{\cal{S}}}

\newcommand{\vars}{\V}
\newcommand{\op}[1]{{\sf #1}}
\newcommand{\true}{1}
\newcommand{\false}{0}
\newcommand{\kw}[1]{{{\,\op{#1}\,}}}
\newcommand{\Not}{\kw{not}}
\newcommand{\Pre}{\kw{\$ init}}
\newcommand{\Sync}{\kw{\mbox{$\hat{~}$}=}}
\newcommand{\Default}{\kw{default}}
\newcommand{\When}{\kw{when}}
\newcommand{\Pred}{{\rm{pred}}}
\newcommand{\Succ}{{\rm{succ}}}
\newcommand{\Par}{{\,|\!\!|\,}}
\newcommand{\In}{{\rm{in}}}
\newcommand{\Out}{{\rm{out}}}
\newcommand{\Def}{{\rm{def}}}
\newcommand{\Use}{{\rm{use}}}

\newcommand{\ra}{\rightarrow}

\newcommand{\Sq}[1]{{[\![{#1}]\!]}}

\newcommand{\sq}[1]{\langle\!\!\langle{#1}\rangle\!\!\rangle}
\newcommand{\idx}[1]{\lceil{#1}\rceil}

\newcommand{\st}{{s.t.}}

\newcommand{\Data}{{\kw{data}}}

\newcommand{\Dataflow}{{\kw{dataflow}}}

\newcommand{\Event}{{\kw{event}}}

\newcommand{\Do}{{\kw{do}}}
\newcommand{\Skip}{{\kw{skip}}}
\newcommand{\End}{{\kw{end}}}
\newcommand{\If}{{\kw{if}}}
\newcommand{\Then}{{\kw{then}}}
\newcommand{\Else}{{\kw{else}}}

\newcommand{\trigger}{{\op{trigger}}}
\newcommand{\reset}{{\op{reset}}}
\newcommand{\block}{{\kw{block}}}

\newcommand{\rra}{\twoheadrightarrow}
\newcommand{\Auto}{{\kw{automaton}}}

\newcommand{\Init}{{\kw{initial}\kw{state}}}

\newcommand{\On}{{\kw{on}}}

\newcommand{\State}{{\kw{state}}}

\newcommand{\dt}[1]{{\rra_{#1}}}
\renewcommand{\st}[1]{{\ra_{#1}}}

\newcommand{\tomodif}[1]{\color{red}{\textbf{A modifier ! }}\color{black}{}}

\begin{document}
\maketitle

{\small
\ \\
L. Besnard, T. Gautier, J. Ouy, J.-P. Talpin \\ INRIA Rennes - Bretagne Atlantique / IRISA, Campus de Beaulieu, F-35042 Rennes Cedex, France \email{\{Loic.Besnard, Thierry.Gautier, Julien.Ouy, Jean-Pierre.Talpin\}@irisa.fr }
\\ \\
J.-P. Bodeveix, A. Cortier, M. Pantel, M. Strecker \\ IRIT-ACADIE, Universit\'e Paul Sabatier, 118 Route de Narbonne, F-31062 Toulouse Cedex 9, France \email{\{bodeveix, cortier, pantel, strecker\}@irit.fr}
\\ \\
G. Garcia \\ Thales Alenia Space, 100 Boulevard Midi, F-06150 Cannes, France \email{gerald.garcia@thalesaleniaspace.com}
\\ \\
A. Rugina \\ EADS Astrium, 31 rue des Cosmonautes, Z.I. du Palays, F-31402 Toulouse Cedex 4, France \email{Ana-Elena.RUGINA@astrium.eads.net}
\\ \\
J. Buisson, F. Dagnat \\ Institut T\'el\'ecom / T\'el\'ecom Bretagne, Technop\^ole Brest Iroise, CS83818, F-29238 Brest Cedex 3, France \email{\{jeremy.buisson, Fabien.Dagnat\}@telecom-bretagne.eu}
}

\ \\
\vspace{-.2cm}
\begin{abstract}
The  SPaCIFY project, which aims at bringing advances in MDE to the satellite flight software industry, advocates a top-down approach built on a domain-specific modeling language named Synoptic. In line with previous approaches to real-time modeling such as Statecharts and Simulink, Synoptic features hierarchical decomposition of application and control modules in synchronous block diagrams and state machines. Its semantics is described in the polychronous model of computation, which is that of the synchronous language \signal.
\end{abstract}

\section{Introduction}
\label{Introduction}

In collaboration with major European manufacturers, the \spacify\ project aims at bringing advances in MDE to the satellite flight software industry. It focuses on software development and maintenance phases of satellite lifecycle. The project advocates a top-down approach built on a Domain-Specific Modeling Language (DSML) named Synoptic. The aim of Synoptic is to support all aspects of embedded flight-software design. As such, Synoptic consists of heterogeneous modeling and programming principles defined in collaboration with the industrial partners and end users of the \spacify\ project.

Used as the central modeling language of the \spacify\ model driven engineering process, Synoptic allows to describe different layers of abstraction: at the highest level, the software architecture models the functional decomposition of the flight software. This is mapped to a dynamic architecture which defines the thread structure of the software. It consists of a set of threads, where each thread is characterized by properties such as its frequency, its priority and its activation pattern (periodic, sporadic).

A mapping establishes a correspondence between the software and the dynamic architecture, by specifying which blocks are executed by which threads. At the lowest level, the hardware architecture permits to define  devices (processors, sensors, actuators, busses) and their properties. 

\begin{figure}[h!]
\begin{center}
\includegraphics[scale=0.7]{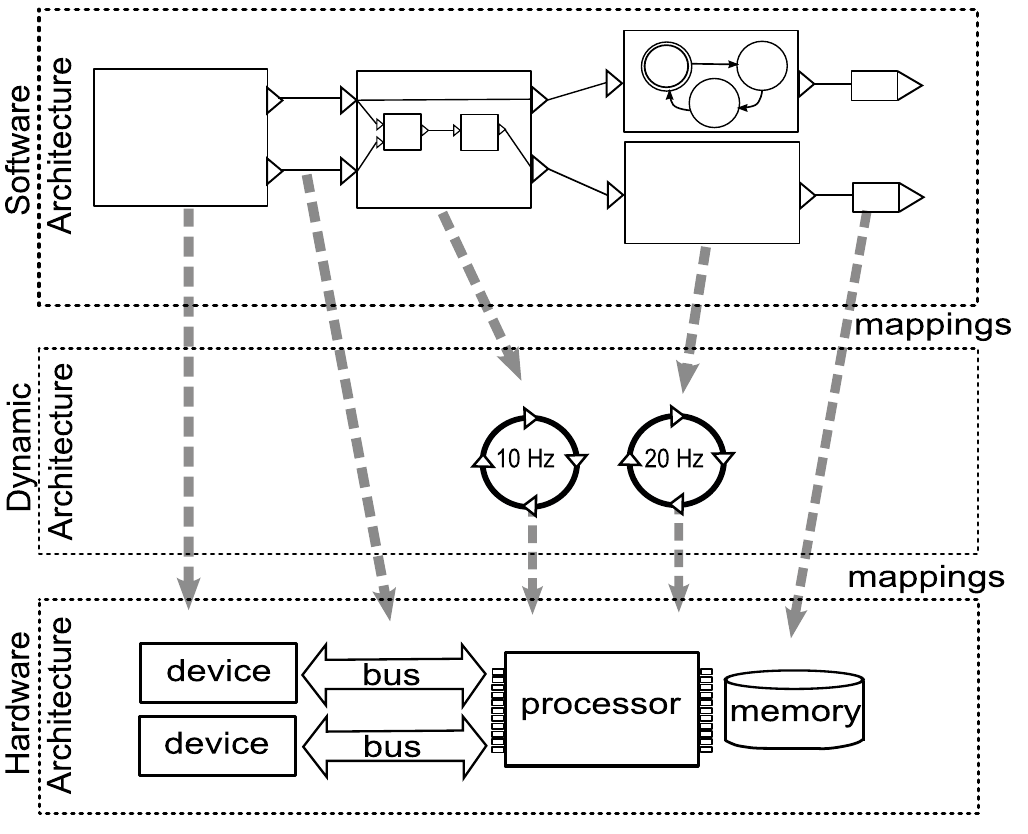}
\end{center}
\caption{Global view: layers and architecture mappings}
\label{mappings}
\end{figure}

Finally, mappings describe the correspondence between the dynamic and hardware architecture on the one hand, by specifying which threads are executed by which processor, and describe a correspondence between the software and hardware architecture on the other hand, by specifying which data is carried by which bus for instance. Figure \ref{mappings} depicts these layers and mappings.

The aim is to synthesize as much of this mapping as possible, for example by appealing to internal or external schedulers. However, to allow for human intervention, it is possible to give a fine-grained mapping, thus overriding or bypassing machine-generated schedules. Anyway, consistency of the resulting dynamic architecture is verified by the \spacify\ tool suite, based on the properties of the software and dynamic model.

At each step of the development process, it is also useful to model  different abstraction levels of the system under design inside a same layer (functional, dynamic or hardware architecture). Synoptic offers this capability by providing an incremental design framework and refinement features.

To summarize, Synoptic deals with data-flow diagrams, mode automata, blocks, components, dynamic and hardware architecture, mapping and timing.

The functional part of the Synoptic language allows to model software architecture. The corresponding sub-language is well adapted to model {\em synchronous islands} and to specify interaction points between these islands and the middleware platform using the concept of {\em external variables}.

Synchronous islands and middleware form a Globally Asynchronous and Locally Synchronous (GALS) system.

\paragraph{Software architecture}

The development of the Synoptic software architecture language has been tightly coordinated with the definition of the GeneAuto language \cite{GeneAuto}. Synoptic  uses essentially two types of modules, called blocks in Synoptic, which can be mutually nested: data-flow diagrams and mode automata. Nesting favors a hierarchical design and enables  viewing the description at different levels of detail.

By embedding blocks in the states of state machines, one can elegantly model operational modes: each state represents a mode, and transitions correspond to mode changes. In each mode, the system may be composed of other sub-blocks or have different connection patterns among components.

Apart from structural and behavioral aspects, the Synoptic software architecture language allows to define temporal properties of blocks. For instance, a block can be parameterized with a frequency and a worst case execution time which are taken into account in the mapping onto the dynamic architecture.

Synoptic is equipped with an assertion language that allows to state desired properties of the model under development. We are mainly interested in  properties that permit to express, for example, coherence of the modes (``if component X is in mode m1, then component Y is in mode m2'' or ``\ldots\ can eventually move into mode m2''). Specific transformations extract these properties and pass them to the verification tools.

The main purpose of this paper is to describe a formal semantics of Synoptic, expressed in terms of the synchronous language \signal\ \cite{Signal,Polychrony}. \signal\ is based on ``synchronized data-flow'' (flows with synchronization): a process is a set of equations on elementary flows describing both data and control. The \signal\ formal model provides the capability to describe systems with several clocks ({\em polychronous} systems) as relational specifications. A brief overview of the abstract syntax of Synoptic is provided in Section~\ref{Syntax}. Then Section~\ref{Semantics} describes the interpretation of each one of these constructions in the model of the \signal\ language.

\section{An overview of Synoptic}
\label{Syntax}

{\em Blocks} are the main structuring elements of Synoptic. A block $\block\,x\,A$ defines a functional unit of compilation and of execution that can be called from many contexts and with different modes in the system under design. A block $x$ encapsulates a functionality $A$ that may consist of sub-blocks, automata and data-flows.  A block $x$ is implicity associated with two signals $x.\trigger$ and $x.\reset$.  The signal $x.\trigger$ starts the execution of $A$.  The specification $A$ may then operate at its own pace until the next $x.\trigger$ is signaled.   The signal $x.\reset$ is delivered to $x$ at some $x.\trigger$ and forces $A$ to reset its state and variables to initial values.
\[
\begin{array}{rl}
(blocks)\quad A,B::= &\block\,x\,A\ |\ \Dataflow\,x\,A\ |\ \Auto\,x\, A\ |\ A\Par B
\end{array}
\]

{\em Data-flows} inter-connect blocks with data and events (e.g.\ trigger and reset signals). A flow can simpliy define a connection from an event $x$ to an event $y$, written $\Event\,x\ra y$, combine data $y$ and $z$ by a simple operation $f$ to form the flow $x$, written $\Data\,y\,f\,z\ra x$ or   feed a signal $y$ back to $x$, written $\Data\,y\Pre v\ra x$.  In a feedback loop, the signal $x$ is initially defined by $x_0=v$. Then, at each occurrence $n>0$ of the signal $y$, it takes its previous value $x_n=y_{n-1}$. The execution of a data-flow is controlled by its parent clock.  A data-flow simultaneously executes each connection it is composed of every time it is triggered by its parent block.
\[
\begin{array}{rl}
(dataflow)\quad A,B::= &\Data\,y\Pre v\ra x\ |\ \Data\,y\,f\,z\ra x\ |\ \Event\,x\ra y\ |\ A\Par B
\end{array}
\]

{\em Actions} are sequences of operations on variables that are performed during the execution of automata. An assignment $x=y\,f\,z$ defines the new value of the variable $x$ from the current values of $y$ and $z$ by the function $f$. The $\Skip$  stores the new values of variables that have been defined before it, so that they become current past it.  The conditional $\If x\Then A\Else B$ executes $A$ if the current value of $x$ is true and executes $B$ otherwise. A sequence $A;B$ executes $A$ and then $B$. 
\[
\begin{array}{rl}
(action)\quad A,B::=& \Skip\ |\ x=y\,f\,z\ |\ \If x\Then A\Else B\ |\ A;B
\end{array}
\]

{\em Automata} schedule the execution of operations and blocks by performing timely guarded transitions. An automaton receives control from its trigger and reset signals $x.\trigger$ and $x.\reset$ as specified by its parent block.  When an automaton is first triggered, or when it is reset, its starts execution from its initial state, specified as $\Init S$. On any state $S:\Do A$, it performs the action $A$. From this state, it may perform an immediate transition to new state $T$, written $S\ra^{\On\,x} T$, if the value of the current variable $x$ is true.  It may also perform a delayed transition to $T$, written $S\rra^{\On\,x} T$, that waits the next trigger before to resume execution (in state $T$). If no transition condition applies, it then waits the next trigger and resumes execution in state $S$.  States and transitions are composed as $A\Par B$. The timed execution of an automaton combines the behavior of an action or a data-flow.  The execution of a delayed transition or of a stutter is controlled by an occurrence of the parent trigger signal (as for a data-flow). The execution of an immediate transition is performed without waiting for a trigger or a reset (as for an action).
\[
\begin{array}{rl}
(automaton)\quad A,B::=& \State S:\Do A\ |\ S\ra^{\On\,x} T\ |\ S\rra^{\On\,x} T\ |\ A\Par B
\end{array}
\]

\section{Polychronous interpretation of Synoptic}
\label{Semantics}

The model of computation on which Synoptic relies is that of the polychronous data-flow language \signal. This section describes how Synoptic programs are interpreted into this core language.

\subsection{A brief introduction to \signal}

In \signal, a process $P$ consists of the composition of simultaneous equations $x=f(y,z)$ over signals $x,y,z$. A delay equation $x= y\Pre v$ defines $x$ every time $y$ is present. Initially, $x$ is defined by the value $v$, and then, it is defined by the previous value of $y$. A sampling equation $x= y\When z$ defines $x$ by $y$ when $z$ is true. Finally, a merge equation $x=y\Default z$ defines $x$ by $y$ when $y$ is present and by $z$ otherwise. An equation $x=y\,f\,z$ can use a boolean or arithmetic operator $f$ to define all of the $n^{th}$ values of the signal $x$ by the result of the application of $f$ to the $n^{th}$ values of the signals $y$ and $z$. The synchronous composition of processes $P\Par Q$ consists of the simultaneous solution of the equations in $P$ and in $Q$. It is commutative and associative. The process $P/x$ restricts the signal $x$ to the lexical scope of $P$.
\[
P,Q ::= x=y\,f\,z\ |\ P/x\ |\ P\Par Q\qquad\mbox{(process)}
\]
In \signal, the presence of a value along a signal $x$ is an expression noted $\hat{~}x$. It is true when $x$ is present. Otherwise, it is absent. Specific processes and operators are defined in \signal\ to manipulate clocks explicitly. We only use the simplest one, $x\Sync y$, that synchronizes all occurrences of the signals $x$ and $y$.

\subsection{Interpretation of blocks}

The execution of a block is driven by the trigger $t$ of its parent block.  The block resynchronizes with that trigger every time, itself or one of its sub-blocks, makes an explicit reference to time (e.g. a \Skip for an action or a delayed transition $S\rra T$ for an automaton).  Otherwise, the elapse of time is sensed from outside the block, whose operations (e.g., on $c_i$), are perceived as belonging to the same period as within $[t_i, t_{i+1}[$. The interpretation implements this feature by encoding actions and automata using static single assignment. As a result, and from within a block, every non-time-consuming sequence of actions $A;B$ or transitions $A\ra B$ defines the value of all its variables once and defines intermediate ones in the flow of its execution.

\subsection{Interpretation of data-flow}

Data-flows are structurally similar to \signal\ programs and equally combined using synchronous composition. The interpretation $\Sq{A}^{rt}=\sq{P}$ of a data-flow (Fig.~\ref{fig2}) is parameterized by the reset and trigger signals of the parent block and returns a process $P$ (the input term $A$ and the output term $P$ are marked by $\Sq{A}$ and $\sq{P}$ for convenience). A delayed flow $\Data\,y\Pre v\ra x$ initially defines $x$ by the value $v$. It is reset to that value every time the reset signal $r$ occurs. Otherwise, it takes the  previous value of $y$ in time.

%

\begin{figure}[h!]
\hrulefill
\[
\begin{array}{@{}r@{\,}c@{\,}l@{}}
\Sq{\Dataflow\,f\,A}^{r\!t}&=&\sq{\Sq{A}^{r\!t}\Par \left(\prod_{x\in\In(A)} x\Sync t\right)}\\
\Sq{\Data\,y\Pre v\ra x}^{r\!t}&=&\sq{x=(v\When r)\Default (y\Pre v)\Par (x\Sync y)}\\
\Sq{\Data\,y\,f\,z\ra x}^{r\!t}&=&\sq{x=y\,f\,z}\\
\Sq{\Event\,y\ra x}^{r\!t}&=&\sq{x=\When y}\\
\Sq{A\Par B}^{r\!t}&=&\sq{\Sq{A}^{r\!t}\Par\Sq{B}^{r\!t}}
\end{array}
\]
\hrulefill
\caption{Interpretation of data-flow connections}\label{fig2}
\end{figure}

In Fig.~\ref{fig2}, we write $\prod_{i\leq n}P_i$ for a finite product of processes $P_1\Par\dots P_n$. Similarly,
$\bigvee_{i\leq n} e_i$ is a finite merge $e_1\Default\dots e_n$.

A functional flow $\Data\,y\,f\,z\ra x$ defines $x$ by the product of $(y,z)$ by $f$.  An event flow $\Event\,y\ra x$ connects $y$ to define $x$. Particular cases are the operator $?(y)$ to convert an event $y$ to a boolean data and the operator $\hat{~}(y)$ to convert the boolean data $y$ to an event. We write $\In(A)$ and $\Out(A)$ for the input and output signals of a data-flow $A$.

By default, the convention of Synoptic is to synchronize the input signals of a data-flow to the parent trigger. It is however, possible to define alternative policies.
One is to down-sample the input signals at the pace of the trigger.  Another is to adapt or resample them at that trigger.


\subsection{Interpretation of actions}

The execution of an action $A$ starts at an occurrence of its parent trigger and shall end before the next occurrence of that event.  During the execution of an action, one may also  wait and synchronize with this event by issuing a $\Skip$.  A \Skip\ has no behavior but to signal the end of an instant: all the newly computed values of signals are flushed in memory and execution is resumed upon the next parent trigger. Action $x!$ sends the signal $x$ to its environment. Execution may continue within the same symbolic instant unless a second emission is performed: one shall issue a \Skip\ before that. An operation $x=y\,f\,z$ takes the current value of $y$ and $z$ to define the new value of $x$ by the product with $f$. A conditional $\If x \Then A \Else B$ executes $A$ or $B$ depending on the current value of $x$. 

As a result, only one new value of a variable $x$ should at most be defined within an instant delimited by a start and an end or a skip. Therefore, the interpretation of an action consists of its decomposition in static single assignment form.  To this end, we use an environment $E$ to associate each variable with its definition, an expression, and a guard, that locates it (in time).

An action holds an internal state $s$ that stores an integer $n$ denoting the current portion of the actions that is being executed. State $0$ represents the start of the program and each $n>0$ labels a $\Skip$ that materializes a synchronized sequence of actions.

The interpretation $\Sq{A}^{s,m,g,E}=\sq{P}_{n,h,F}$ of an action $A$ (Fig.~\ref{fig3}) takes as parameters the state variable $s$, the state $m$ of the current section, the guard $g$ that leads to it, and the environment $E$. It returns a process $P$, the state $n$ and guard $h$ of its continuation, and an updated environment $F$. We write $\Use^g_E(x)$ for the expression that returns the definition of the variable $x$ at the guard $g$ and $\Def^g_E(x)$ for storing the final values of all variables $x$ defined in $E$ (i.e., $x\in\vars(E)$) at the guard $g$.
\[
\begin{array}{@{}r@{\,}c@{\,}l@{}}
\Use^g_E(x)&=&if\,x\in\vars(E)\,then\,\sq{E(x)}\,else\,\sq{(x\Pre 0)\When g}\\
\Def_{g}(E)&=&\prod_{x\in\vars(E)}\left(x=\Use^g_E(x)\right)
\end{array}
\]
Execution is started with $s=0$ upon receipt of a trigger $t$. It is also resumed from a skip  at $s=n$ with a trigger $t$. Hence the signal $t$ is synchronized to the state $s$ of the action.  The signal $r$ is used to inform the parent block (an automaton) that the execution of the action has finished (it is back to its initial state $0$).  An \End\  resets $s$ to $0$, stores all variables $x$ defined in $E$ with an equation $x=\Use^g_E(x)$ and finally stops (its returned guard is \false). 
A \Skip\  advances $s$ to the next label $n+1$ when it receives control upon the guard $e$ and flushes the variables defined so far. It returns a new guard $(s\Pre 0)=n+1$ to resume the actions past it. An action $x!$ emits $x$ when its guard $e$ is true. A sequence $A;B$ evaluates $A$ to the process $P$ and passes its state $n_A$, guard $g_A$, environment $E_A$ to $B$. It returns $P\Par Q$ with the state, guard and environment of $B$.
Similarly, a conditional evaluates $A$ with the guard ${g\When x}$ to $P$ and $B$ with ${g\When\Not x}$ to $Q$. It returns $P\Par Q$ but with the guard $g_A\Default g_B$. All variables $x\in X$, defined in both $E_A$ and $E_B$, are merged in the environment $F$.

\begin{figure}[h!]
\hrulefill
\[
\begin{array}{@{}r@{\,}c@{\,}l@{}}
\Sq{\Do A}^{r\!t}&=&\sq{\left(P\Par s \Sync t\Par r=(s=0)\right)/s}\,
where\,\sq{P}_{n,h,F}=\Sq{A;\End}^{s,0,\left((s\,\op{pre}\,0)=0\right),\emptyset}\\[+\smallskipamount]
\Sq{\End}^{s,n,g,E}&=&\sq{s=0\When g\Par\Def_{g}(E)}_{0,\false,\emptyset}\\
\Sq{\Skip}^{s,n,g,E}&=&\sq{s=n+1\When g\Par\Def_{g}(E)}_{n+1,\left((s\,\op{pre}\,0)=n+1\right),\emptyset}\\
\Sq{x!}^{s,n,g,E}&=&\sq{x=\true\When g}_{n,g,E}\\
\Sq{x=y\,f\,z}^{s,n,g,E}&=&\sq{x=e}_{n,g,E_x\uplus\{x\mapsto e\}}\,
where\,e=\sq{f(\Use^g_E(y),\Use^g_E(z))\When g}\\[+\smallskipamount]
\Sq{A;B}^{s,n,g,E}&=&\sq{P\Par Q}_{n_B,g_B,E_B}\,
where\,\sq{P}_{n_A,g_A,E_A}=\Sq{A}^{s,n,g,E}\,
and\,\sq{Q}_{n_B,g_B,E_B}=\Sq{B}^{s,n_A,g_A,E_A}\\
\end{array}
\]
\[
\begin{array}{@{}r@{\,}c@{\,}l@{}}
\Sq{\If\,x\,\Then\,A\,\Else\,B}^{s,n,g,E}&=&\sq{P\Par Q}_{n_B,(g_A\Default g_B),(E_A\uplus E_B)}\\
where\,\sq{P}_{n_A,g_A,E_A}&=&\Sq{A}^{s,n,(g\,\op{when}\,\Use^g_E(x)),E}\,
and\,\sq{Q}_{n_B,g_B,E_B}=\Sq{B}^{s,n_A,(g\,\op{when}\,\op{not}\,\Use^g_E(x)),E}\\
\end{array}
\]
\hrulefill
\caption{Interpretation of timed sequential actions}\label{fig3}
\end{figure}

In Fig.~\ref{fig3}, we write $E\uplus F$ to merge the definitions in the environments $E$ and $F$. For all variables $x\in\vars(E)\cup\vars(F)$ in the domains of $E$ and $F$,
\[
(E\uplus F)(x)=\left\{\begin{array}{ll}
E(x),&x\in\vars(E)\setminus\vars(F)\\
F(x),&x\in\vars(F)\setminus\vars(E)\\
E(x)\Default F(x),&x\in\vars(E)\cap\vars(F)\\
\end{array}\right.
\]
Note that an action cannot be reset from the parent clock because it is not synchronized to it. A sequence of emissions $x!;x!$ yields only one event along the signal $x$ because they occur at the same (logical) time, as opposed to $x!;\Skip;x!$ which sends the second one during the next trigger.

\subsection{Interpretation of automata}

An automaton describes a hierarchic structure consisting of actions that are executed upon entry in a state by immediate and delayed transitions. An immediate transition occurs during the period of time allocated to a trigger. Hence, it does not synchronize to it. Conversely, a delayed transition occurs upon synchronization with the next occurrence of the parent trigger event. As a result, an automaton is partitioned in regions.  Each region corresponds to the amount of calculation that can be performed within the period of a trigger, starting from a given initial state.

\paragraph*{Notations}

We write $\st A$ and $\dt{A}$ for the immediate and delayed transition relations of an automaton $A$.  We write $\Pred_{\st A}(S)=\{T\,|\,(T,x,S)\in R\}$ and $\Succ_{\st A}(S)=\{T\,|\,(S,x,T)\in R\}$ (resp.\ $\Pred_{\dt A}(S)$ and $\Succ_{\dt A}(S)$) for the predecessor and successor states of the immediate (resp.\ delayed) transitions $\st A$ (resp.\ $\dt{A}$) from a state $S$ in an automaton $A$.Finally, we write $\vec S$ for the region of a state $S$. It is defined by an equivalence relation.
\[
\forall S,T\in\S(A),\,\left((S,x,T)\in\st A\right)\Leftrightarrow \vec S = \vec T
\]
For any state $S$ of $A$, written $S\in\S(A)$, it is required that the restriction of $\st A$ to the region $\vec S$ is acyclic. Notice that, still, a delayed transition may take place between two states of the same region.

\paragraph*{Interpretation}

An automaton $A$ is interpreted by a process $\Sq{\Auto x\,A}^{r\!t}$ parameterized by its parent trigger and reset signals. The interpretation of $A$ defines a local state $s$. It is synchronized to the parent trigger $t$.  It is set to $0$, the initial state, upon receipt of a reset signal $r$ and, otherwise, takes the previous value of $s'$, that denotes the next state. The interpretation of all states is performed concurrently.

We give all states $S_i$ of an automaton $A$ a unique integer label $i=\idx{S_i}$ and designate with $\idx{A}$ its number of states. $S_0$ is the initial state and, for each state of index $i$, we call $A_i$ its action $i$ and $x_{ij}$ the guard of an immediate or delayed transition from $S_i$ to $S_j$.
\[
\begin{array}{l}
\Sq{\Auto x\,A}^{r\!t}=\\\qquad\sq{\left(
t\Sync s\Par s=(0\When r)\Default(s'\Pre 0)\Par
\left(\prod_{S_i\in \S(A)}\Sq{S_i}^{s}\right)\right)/ss'}
\end{array}
\]
The  interpretation $\Sq{S_i}^{s}$ of all states $0\leq i<\idx{A}$ of an automaton (Fig.~\ref{fig6}) is implemented by a series of mutually recursive equations that define the meaning of each state $S_i$ depending on the result obtained for its predecessors $S_j$ in the same region.  Since a region is by definition acyclic, this system of equations has therefore a unique solution.

The interpretation of state $S_i$ starts with that of its actions $A_i$. An action $A_i$ defines a local state $s_i$ synchronized to the parent state $s=i$ of the automaton. The automaton stutters with $s'=s$ if the evaluation of the action is not finished: it is in a local state $s_i\neq 0$. 

Interpreting the actions $A_i$ requires the definition of a guard $g_i$ and of an environment $E_i$. The guard $g_i$ defines when $A_i$ starts. It requires the local state to be $0$ or the state $S_i$ to receive control from a predecessor $S_j$ in the same region (with the guard $x_{ji}$).

The environment $E_i$ is constructed by merging these $F_j$ returned by its immediate predecessors $S_j$.  Once these parameters are defined, the interpretation of $A_i$ returns a process $P_i$ together with an exit guard $h_i$ and an environment $F_i$ holding the value of all variables it defines.

Upon evaluation of $A_i$, delayed transition from $S_i$ are checked.  This is done by the definition of a process $Q_i$ which, first, checks if the guard $x_{ij}$ of a delayed transition from $S_i$ evaluates to true with $F_i$.  If so, variables defined in $F_i$ are stored with $\Def_{h_i}(F_i)$.

All delayed transitions from $S_i$ to $S_j$ are guarded by $h_i$ (one must have finished evaluating $i$ before moving to $j$) and a condition $g_{ij}$, defined by the value of the guard $x_{ij}$.  The default condition is to stay in the current state $s$ while $s_i\neq 0$ (i.e. until mode $i$ is terminated). 

Hence, the next state from $i$ is defined by the equation $s'=s'_i$. The next state equation of each state is composed with the other to form the product $\prod_{i< \idx{A}}s'=s'_i$ that is merged as $s'=\bigvee_{i< \idx{A}}s'_i$.

\begin{figure}[h!]
\hrulefill
\[
\begin{array}[t]{@{}@{}l}
\forall i< \idx{A},\,\Sq{S_i}^{s}=
\left(P_i\Par Q_i\Par s_i\Sync\When (s=i)\Par s'=s'_i\right)/s_i\,where\\[+\medskipamount]
\quad \sq{P_i}_{n,h_i,F_i}=\Sq{A_i}^{s_i,0,g_i,E_i}\\
\quad Q_i=\prod_{(S_i,x_{ij},S_j)\in\dt A}\left(\Def_{h_i\,\op{when}\,(\Use_{F_i}(x_{ij}))}(F_i)\right)\\[+\smallskipamount]
\quad E_i=\,\biguplus_{S_j\in\Pred_{\st A}(S_i)} F_j\\[+\smallskipamount]
\quad g_i=\true\When(s_i\Pre 0=0)\Default\left(\bigvee_{(S_j,x_{ji},S_i)\in\st A}(\Use_E(x_{ji}))\right)\\[+\smallskipamount]
\quad g_{ij}=h_i\When(\Use_{F_i}(x_{ij})),\,\forall (S_i,x_{ij},S_j)\in\dt A\\[+\smallskipamount]
\quad s'_i=(s\When s_i\neq 0)\Default\!\left(\bigvee_{(S_i,x_{ij},S_j)\in\dt A}(j\When g_{ij})\right)
\end{array}
\]
\hrulefill
\caption{Recursive interpretation of a mode automaton}\label{fig6}
\end{figure}




\section{Conclusion}
\label{Conclusion}

Synoptic has a formal semantics, defined in terms of the synchronous language \signal. On the one hand, this allows for neat integration of verification environments for ascertaining properties of the system under development. On the other hand, a formal semantics makes it possible to encode the meta-model in a proof assistant. In this sense, Synoptic will profit from the formal correctness proof and subsequent certification of a code generator that is  under way in the GeneAuto project.
Moreover, the formal model of \signal\ is the basis for the Eclipse-based polychronous modeling environment SME~\cite{Polychrony,SME}. SME is used to transform Synoptic diagrams and generate executable C code.

\bibliographystyle{eptcs} 

{\small
\begin{thebibliography}{1}

\bibitem{GeneAuto} A. Toom, T. Naks, M. Pantel, M. Gandriau and I. Wati: 
\newblock \emph{GeneAuto: An Automatic Code Generator for a safe subset of 
SimuLink/StateFlow}.
\newblock {\sl European Congress on Embedded Real Time Software (ERTS'08)},
Soci\'et\'e des Ing\'enieurs de l'Automobile, (2008).

\bibitem{Signal} P. Le Guernic, J.-P. Talpin and J.-C. Le Lann:
\newblock \emph{Polychrony for system design}.
\newblock {\sl Journal for Circuits, Systems and Computers},
Special Issue on Application Specific Hardware Design,
World Scientific, (2003). 

\bibitem{Polychrony} Polychrony and SME. 
\newblock Available at \url{http://www.irisa.fr/espresso/Polychrony}. 

\bibitem{SME} C. Brunette, J.-P. Talpin, A. Gamati\'e and T. Gautier:
\newblock \emph{A metamodel for the design of polychronous systems}.
\newblock {\sl The Journal of Logic and Algebraic Programming}, 78,
Elsevier, (2009). 

\end{thebibliography}
}

\end{document}